\documentclass[sigconf,natbib=false,nonacm]{acmart}
\setcopyright{none}

\AtBeginDocument{%
  }

\RequirePackage[
  datamodel=acmdatamodel,
  style=acmnumeric,
  backend=biber,
  sorting=none
  ]{biblatex}

\usepackage{svg}

\begin{document}

\title{DPU or GPU for Accelerating Neural Networks Inference – \\
Why not both? Split CNN Inference}

\author{Ali Emre Oztas, Mahir Demir, James Garside, and Mikel Luj\'an}
\affiliation{%
  \institution{The University of Manchester}
  \country{United Kingdom}
}




\renewcommand{\shortauthors}{Oztas et al.}

\begin{abstract}
  Video and image streaming on edge devices requires low latency. To address this, Neural Networks (NNs) are widely used, and prior work mainly focuses on accelerating them with single hardware units such as Graphics Processing Units (GPUs), Field Programmable Gate Arrays (FPGAs), and Deep Learning Processing Units (DPUs). However, further reductions in latency can be observed by combining these units.
  
In this paper, partitioning CNN inference across DPU and GPU (Split CNN Inference) is proposed. The first partition runs on the AI engines (DPU) of a Versal VCK190, which consists of initial CNN layers processing the input images. The DPU processes the first partition near the source of the data. Pipelined asynchronously, a GPU runs the remaining layers. The GPU (NVIDIA RTX 2080) processes the second partition, albeit having reduced the data transfer between the data source (storage/camera) and the GPU. Furthermore, a Graph Neural Network (GNN)–based partition index prediction method is proposed to automate the partitioning of CNNs needed for Split Inference.

Well established models such as LeNet-5, ResNet18/50/101/152, VGG16, and MobileNetv2 are analyzed. Results demonstrate up to 2.48× latency improvement over DPU-only execution and up to 3.37× over GPU-only execution. The trained GNN model splits the layers between the appropriate devices with 96.27\% accuracy.
\end{abstract}



\maketitle

\section{Introduction}
Since the term "pixel" was coined in 1965, image processing, computer graphics, and computer vision have continuously evolved, enabling real-time streaming applications with up to 8K image quality. In those 6 decades,  many codecs, standards, hardware, and increasingly Machine Learning (ML) models have been deployed. This paper considers one of the most prevalent ML models for such applications, Convolutional Neural Networks (CNNs). 

Since the initial successes of the LeNet family of CNNs, in particular LeNet-5 in 1998 \cite{lenet}, and after reaching an inflection point with AlexNet in 2012, with its performance in the ImageNet Challenge \cite{img_data, img_correct}, CNNs have become indispensable in image processing for multiple problems, such as recognizing handwritten digits, object detection, object classification, or image segmentation.

\begin{figure*}[tb!]
  \centering
  \includegraphics[width=1.35\columnwidth]{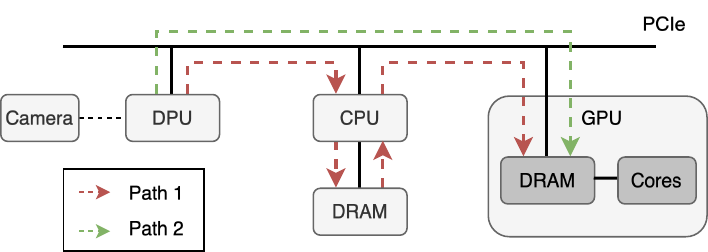}
  \caption{
   Example of an edge system for image processing with DPU and GPU.}
  \label{main}
\end{figure*}

When considering image processing on edge systems using CNNs, it is widely accepted that they need to take advantage of hardware accelerators, specialized hardware units, to be able to operate at demanding frame rates. Examples can be found in \cite{mul} for detecting humans within bounding boxes, and in \cite{gun, edge} for object detection, or in \cite{real} for enhancing from low resolution to high resolution. 

Hardware accelerators, such as GPUs, can speed up computationally intensive tasks by enabling the offloading of computation from general-purpose processing cores. GPUs play an important role in CNN workloads because of their parallelizable nature and predictable memory access patterns. Hardware accelerators can take the form of GPUs, FPGAs, or programmable neural accelerators (e.g., Google’s Tensor Processing Unit (TPU) has produced 6 different iterations over time \cite{tpu, google}). Each hardware accelerator family has their own benefits and shortcomings, and thus forms of programmable neural accelerators (e.g., TPUs) are appearing in GPUs, general-purpose processing cores, and FPGAs.

Nonetheless, modern FPGAs have the advantage of having direct interfaces to cameras (such as MIPI). Furthermore, recent AMD FPGA devices include DPUs as a type of neural accelerator. For example, the DPU of Versal VCK190 FPGA includes AI engine tiles containing an array of Very Long Instruction Word (VLIW), Single Instruction Multiple Data (SIMD) processing engines, and memories \cite{versal}. Figure \ref{main} presents the kind of edge system considered in this paper, where the DPU is a Versal, and the GPU is an NVIDIA RTX 2080. The DPU can process data near the source of information (near sensor/near storage), while the GPU can process the outcome of the DPU.

\begin{figure}[tbh!]
  \centering
  \includegraphics[width=\linewidth]{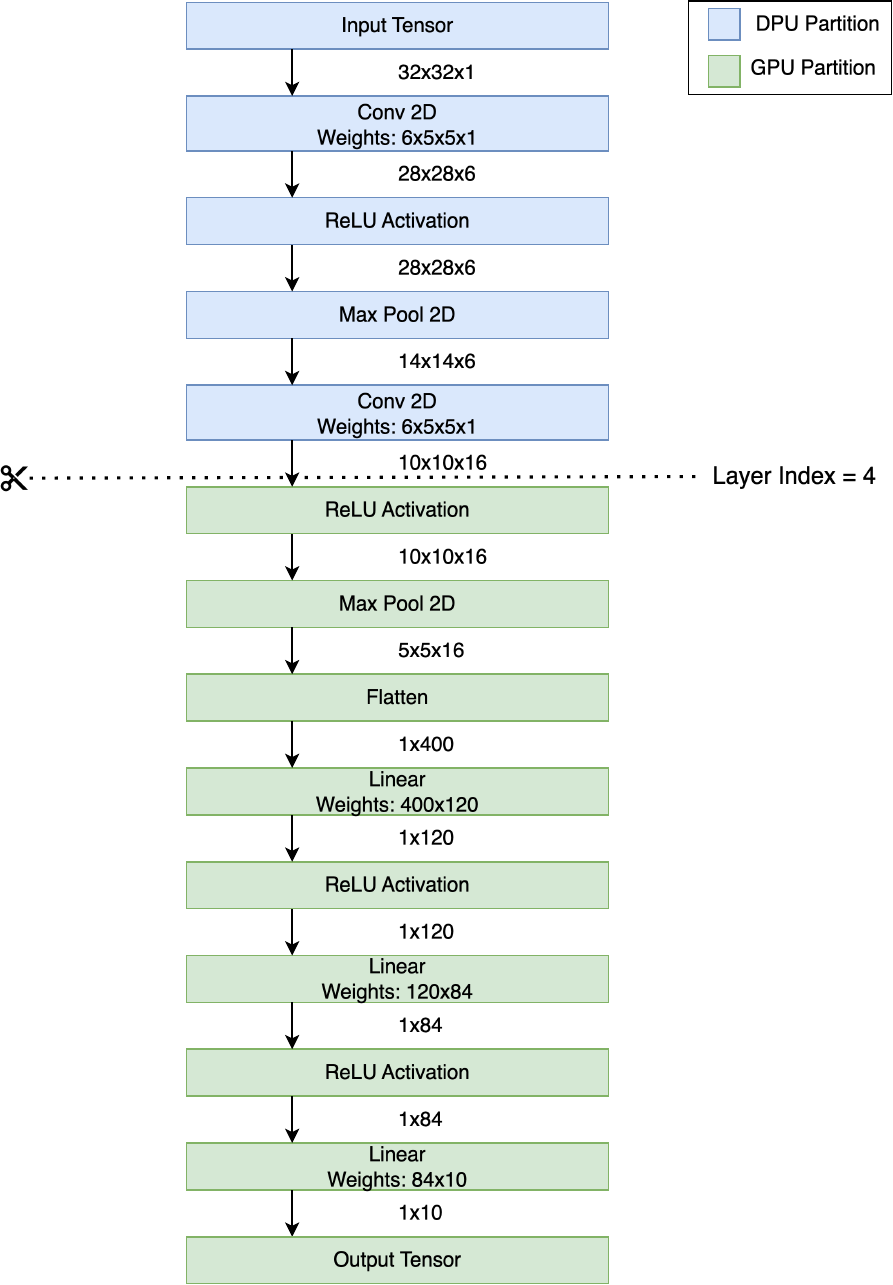}
  \caption{
  Example of optimal partitioning of LeNet-5 inference across DPU and GPU.}
  \label{lenet_part}
  
\end{figure}

Using only one type of hardware accelerator for the entire CNN inference may not be optimal. Different parts of the inference may have different computational patterns, and thus, it may be more efficient to offload specific parts to different accelerators. However, there is limited published research on combining different types of processing units for inference. Examples involving combinations of FPGA, DPU, general processing cores, or GPU to accelerate deep neural network models can be found in \cite{secda,ago,kres,an,yjang,smart}, with  \cite{secda,kres,ago, an} focusing on CNN inference.

When different types of processing elements are used for inference, the next question would be how to efficiently partition the model computation between these processing elements. Whereas an exhaustive search yields the theoretical optimum point, it is not always computationally feasible. A myriad of approaches requiring less computation have appeared in the literature. Some examples can be given as heuristics, evolutionary algorithms, or neural network solutions \cite{kres,zhou,ago}.

This paper analyzes partitioning CNN inference across DPU and GPU, hereafter \textit{Split CNN Inference}. The subset of CNN layers that processes the input data (images) constitutes the first partition. This partition is allocated to the DPU. The remaining subset of CNN layers constitutes the second partition and is allocated to the GPU. The GPU-CNN partition takes the output from the DPU-CNN partition as input data. Thus, the data communicated between the two CNN partitions has to be transferred from the DPU to the GPU. Figure \ref{lenet_part} presents the optimal Split CNN Inference for LeNet-5, where, despite having a limited number of layers, it is still possible to obtain small reductions in latency (See Section \ref{sec:experiments}). 

We run the partitioned CNN layers across the DPU and GPU in a pipelined fashion. Afterwards, we calculate the speedup for each partition. By selecting the partition point with maximum speedup, we ensure that the selected partition is optimal for the given hardware configuration. That makes our partitioning method hardware-aware. This theoretical calculation is made with an exhaustive search. Furthermore, to make it more user-friendly, we also propose a GNN-based prediction model to find partition indices. This approach provides near-optimal results by trading off computation time with accuracy.

The main contributions of this paper can be summarized as:
\begin{itemize}
    \item A fine-grained CNN partitioning approach across DPU and GPU accelerators for improving inference performance in streaming applications.
    \item A novel systematic analysis of pipeline partitioning for CNN inference across DPU and GPU, providing insights into performance trade-offs.
    \item Split CNN inference is evaluated with well-known CNN models, and the results show up to 2.48x latency improvement over DPU-only execution and up to 3.37x over GPU-only execution, demonstrating the effectiveness of Split CNN Inference.
    \item  A novel GNN-based model to predict the partition index for Split CNN Inference. Relative to the optimal partition indices, this GNN model has 96.27\% accuracy.
\end{itemize}


\section{Related Work}
\label{sec:related}

CNNs are attractive for partitioning because there is no specific computing unit architecture that is the most efficient for all layer types. Previous work has explored splitting CNN across CPU/GPU and CPU/FPGA. For example, in \cite{yjang}, a three-way partitioning scheme for training (not inference) is described over storage devices, DPUs, and host general processing cores. They demonstrate a 12.2\% to 31.0\% reduction in training time for vision and recommendation-system tasks. Their optimal partition index is an exhaustive search to find the optimal partition. 
An \textit{et al}.\ \cite{an} explore offloading only the first convolutional and pooling layers of CNNs to a near-storage FPGA accelerator. Thus, they do not consider the general partitioning of CNNs. Haris \textit{et al}.\ \cite{secda} introduce a hardware/software co-design methodology (SECDA-TFLite) for designing NN accelerators for resource-constrained edge devices, which contain FPGAs (Pynq). As the full NNs do not fit on the accelerator in the FPGA, SECDA-TFLite creates the software ''glue" for data communication and multi-threading so that the convolutional layers execute on the accelerator being designed and all other layers on the low-power ARM CPUs; the partition is driven by the small size of the FPGA and is not considered in general. Kreß \textit{et al}.\ \cite{kres} propose a theoretical analysis and partitioning framework for DNN inference, tailored for distributed ASIC systems, which are represented as a graph. This graph is then partitioned with an evolutionary algorithm, and needs to be rerun for each different NN model. For big.LITTLE systems, Aghapour \textit{et al}.\ \cite{ago} propose a layer-switching architecture between the ARM CPU and GPU to reduce latency in CNN inference. Despite the switching overheads, the total latency is reduced on average by 4.72\%. To partition, they use exhaustive search to get all latency results for a given CNN model, and then a greedy search to select the feasible processing unit for each layer in single-image inference. \cite{geometric} describes a method to partition operations within a layer. Their method minimizes the latency and energy consumption of a deep learning model using geometric programming by iteratively finding the optimal partition, taking a linear amount of time. Such an approach is not suitable for streaming scenarios.
The aforementioned works either use a full exhaustive search to find latencies or use computationally intensive algorithms to predict the partitioning index. 

Recently, there have been works that demonstrate the efficiency of GNNs regarding performance prediction and computation mapping. Chai \textit{et al}.\ \cite{chai} used their GNN model called PerfSAGE to predict latency, energy, and memory usage of arbitrary DNNs that run on edge devices. They achieved the state-of-the-art \textless5 Mean Absolute Percentage Error. Zhou \textit{et al}.\ \cite{zhou} proposed the first automatic framework that co-designs the architecture search and the mapping of each operation on Device-Edge hierarchies for GNN models. To find the optimal architecture, they predicted system latency using a GNN-based predictor. However, their model takes specific device-edge pairs as input, which limits its scalability.

To sum up, there is evidence that it is viable to split up deep learning models across different processing units to improve performance. Split CNN Inference reduces latency using DPU and GPU, taking advantage of the proximity of DPU to the image source. Moreover, none of the prior work has performed a systematic analysis of partitioning CNN inference across DPU and GPU. Finally, our work provides an automatic CNN partitioning approach using GNNs, with close to optimal results.

\begin{figure}[tbh!]
  \centering
  \includegraphics[width=1\columnwidth]{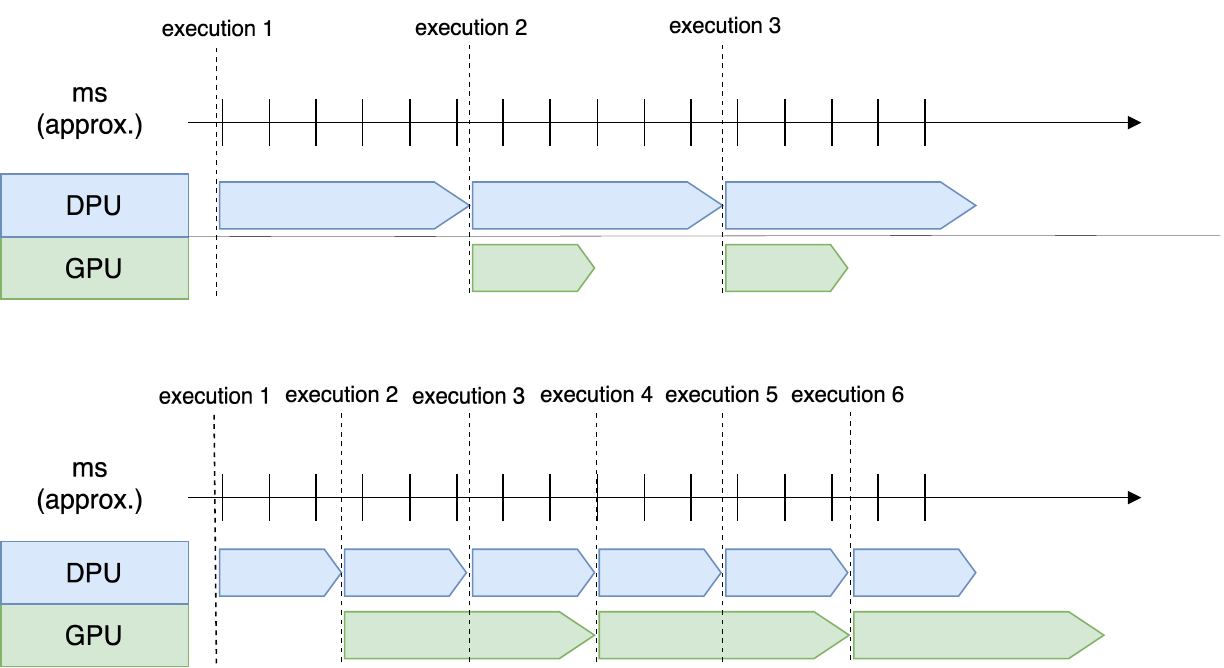}
  \caption{Illustration of increasing parallelism and reducing latency with Split CNN inference.}
  \label{time}
\end{figure}

\section{Split CNN Inference}
\subsection{Partitioning CNN Models}
\label{sec:splitCNN}

Split CNN Inference accelerates inference by partitioning the CNN layers and pipelining them across the DPU and GPU in a hardware-aware manner. Initially, a CNN model is partitioned by searching for the optimal split point (layer) that yields the minimum latency given the DPU and GPU characteristics. The latency term refers to the latency of performing inference on a large number of images. Then, the first partition of the model runs on the DPU of the Versal FPGA. AMD software (Vitis AI) creates designs for the FPGA/DPU, which outputs an INT8 precision output tensor and a 4-byte scale variable to dequantize the output tensor. Later, the output tensor is sent to the GPU along with the scale variable.

The destination of the output tensor can be one of two options depending on the configuration and PCIe capabilities (see Figure \ref{main}). For the first configuration (Path 1), data is moved to the DRAM of the host prior to being moved to the DRAM of the GPU. For the second configuration (Path 2), it is written directly to the DRAM of the GPU (e.g., \ PCIe 5 supports such peer-to-peer communication), optimizing data transfer latency. For both of the paths, we use a queue data structure to store the output tensors.

After sending the output tensor, the DPU can start processing the next image. The NVIDIA RTX
2080 GPU has hardware support for 16-bit precision floating-point arithmetic. Newer NVIDIA/AMD GPUs would enable us to use 8-bit/4-bit floating-point arithmetic. Thus, the execution of the DPU partition and output tensor transfer can be overlapped with GPU partition execution to improve parallelism. Figure \ref{time} shows timing diagrams and illustrates the latency reduction.

\begin{figure*}[bth!]
  \centering
  \includegraphics[width=1.9\columnwidth]{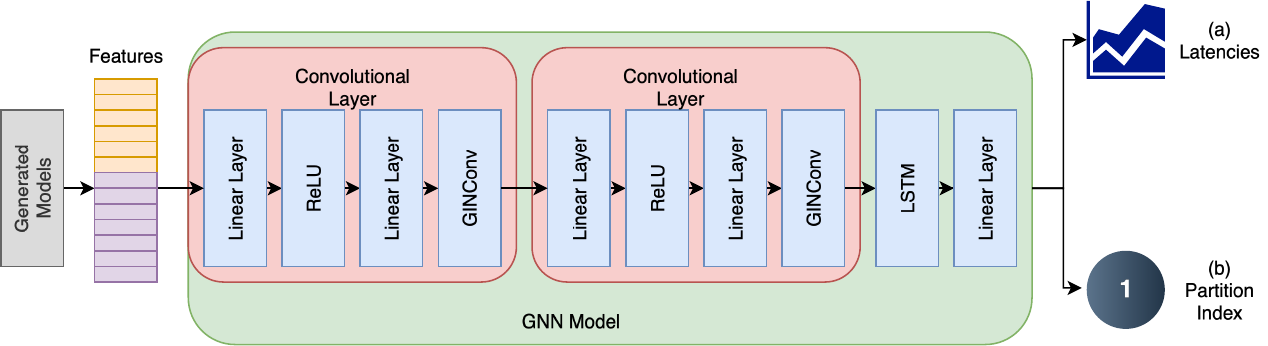}
  \caption{
   Illustration of proposed GNN models to (a) predict model inference latencies or (b) partition index.}
  \label{model}
\end{figure*}

\subsection{Predicting Where to Partition}
\label{sec:algorithm}

The most accurate way to find the optimal partition is by exhaustive search. However, this requires one CNN model inference per partition index, making it infeasible due to the high time complexity. As such, it is necessary to have a time-bound approach to find optimal points. Heuristics and machine-learning models can be applied in this context.

The problem statement can be formulated in two different ways. The first way is that, given the layer properties of a CNN model, such as input size, kernel size, channel size, etc., there is an algorithm that fits a curve where the x-axis is the layer index and the y-axis is the execution time for running layers before or after the layer index. After execution on the two devices, partition-wise latency profiles can be generated and combined with DPU-to-GPU data-transfer latencies to estimate the optimal partition point, following the measurement methodology described in Section \ref{sec:experiments}. The second formulation uses the same features but directly predicts which layer is the best partitioning point. Both of these formulations can be addressed with Machine Learning regression models.

We have tried different algorithms to solve both problem representations. Initially, we tried low-complexity tree-based models with the help of AutoGluon \cite{autogluon_tab} (a state-of-the-art Auto ML system), for both regression versions. However, in both cases, we got inaccurate results. The reason is that AutoGluon models do not capture the sequential correlation between the CNN layers. 

Based on this observation, we represent each CNN model as a graph, although still considering the two problem formulations. The layers are the nodes of the graph, and the edges are the connections of the graph. The node features are similar to those used in \cite{chai}, with the additional inclusion of size features for input and output tensors. The important point for edges is that, for the DPU part, the connected edges follow the dataflow direction, and in the GPU part, the connected edges follow the reverse to preserve causality. The latency of running the first partition depends upon layers before the partition index, and vice versa for the second partition. For edges, we do not have any features, unlike \cite{chai}.

To predict latencies for a given graph, we train a GNN model. The first six layers of the GNN model are Graph Convolution (GCN) and Linear layers to pass information between nodes. The GNN has a Long Short Term Memory (LSTM) \cite{lstm} layer followed by a Linear layer to learn accumulated latency up to that node. Utilizing an LSTM layer significantly improves accuracy as it captures temporal features. The LSTM follows the forward direction for the DPU model and the reverse direction for the GPU model. Figure \ref{model} depicts the full GNN model for both problem representations.

\section{Experiments and Discussion}
\label{sec:experiments}
\subsection{Experimental Setup and Latency Measurement}
\label{sec:meas}
To measure the latency of Split CNN Inference, we divide the selected CNN models into two separate partitions. The first partition runs on the DPU of the Versal VCK190 FPGA board. This DPU is a C32B6CU1L2S2 system of DPUCVDX8G \cite{dpu} and is composed of a hard-core AI engine and PL parts. The second partition runs on an NVIDIA RTX 2080 GPU. The host CPU is responsible for running the host code that orchestrates data transfers. The host CPU is an Intel i9-7900X for the experiments.

The experiments use ResNet18, ResNet50, ResNet101, ResNet152, VGG16, and MobileNetV2 models covering both residual and non-residual models \cite{res,vgg, mobile}. These models are included in model zoos of different deep learning frameworks \cite{bench, zoo, keras, table}, providing independently heavily optimized baselines. The experiments also include LeNet (although not a state-of-the-art CNN anymore) to continue with the running example and to illustrate a very pessimistic situation with little chance of reducing inference latency.

Residual models are CNN models that include shortcut connections from previous layers to later ones. For example, the ResNet family is a residual model with basic block granularity, where the basic block is the part residing between shortcut connections. We did not split basic blocks to reduce transferred data and system complexity. 

Similar to ResNet models, MobileNetv2 has inverted residual layers, which we did not split further. In VGG16, we used layer granularity for splitting because it does not include any residual connections. As a result of this splitting strategy, ResNet18, ResNet50, ResNet101, ResNet152, MobileNetV2, and VGG16 have 16, 24, 41, 57, 24, and 41 different splits, respectively, where the first split is the entire model on the GPU and the last split is the entire model on the DPU. All of the first partitions of these splits are quantized to 8-bit precision and compiled using Vitis-AI software. Then, the accuracy of the models is measured on a 50K subset of the ImageNet 2012 dataset \cite{img_data, img_correct}. For clarity, accuracy results are not included in the latency plot in Fig.~\ref{vgg}; however, classification accuracy was verified after quantization, and only minor degradation was observed, e.g., approximately a 1\% drop for VGG16.

\begin{figure*}[tbh!]
  \centering
  \includegraphics[width=0.8\linewidth]{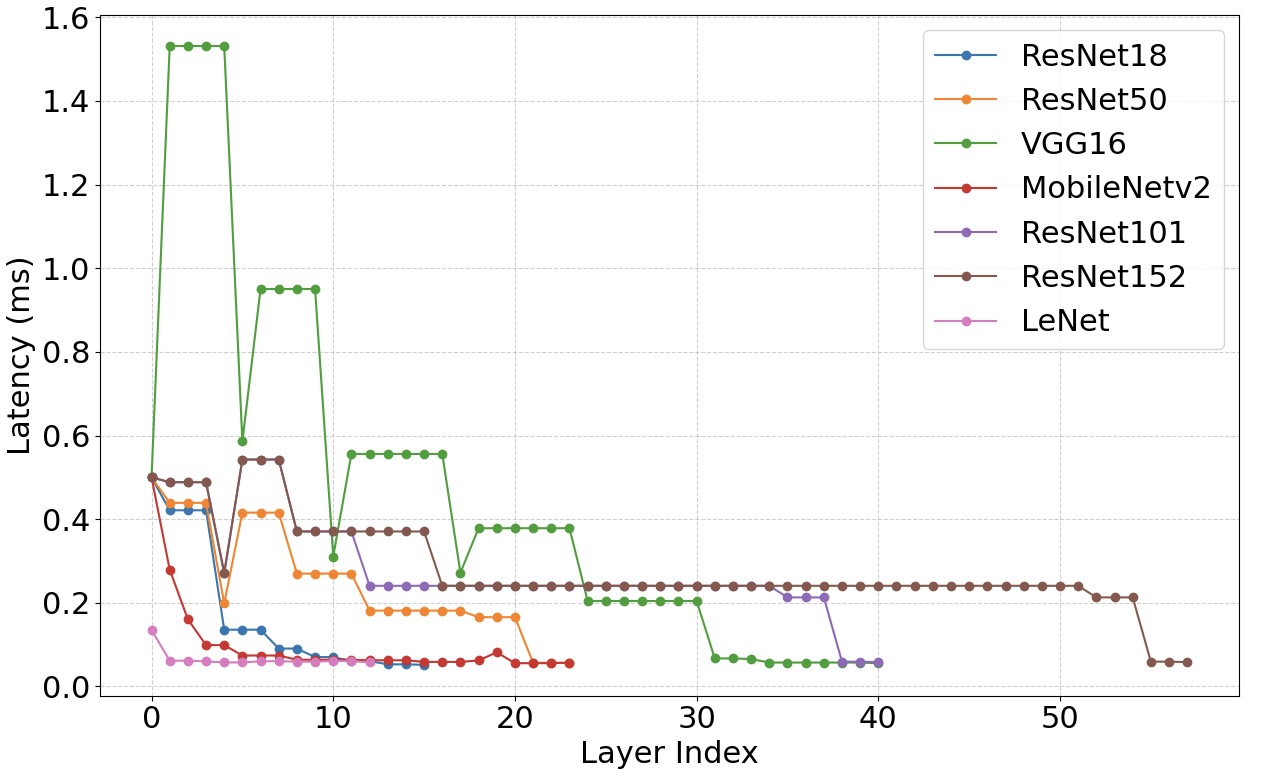}
  \caption{Exploration of data transfer latency between DPU and host DRAM across partitions.}
  \label{latent}
\end{figure*}

The experiments capture (1) the runtime of the first partition execution, (2) PCIe data transfer times between DPU and DRAM using sizes of output tensors (see Figure \ref{latent}). All these tensors are of 8-bit precision. The data is transferred over PCIe Gen3x4, which has 8GB/s bandwidth. The second partition (GPU) uses 16-bit floating-point precision and non-quantized models. To capture different PCIe versions/capabilities, as illustrated in Figure \ref{main}, there are two paths. Following Path 2 (peer-to-peer PCIe), the experiments measure the execution time after moving tensors to the GPU DRAM from the host DRAM. Following Path 1, we also include host DRAM-to-GPU data transfer.



\begin{table}[tbh!] 
    \centering
    \renewcommand{\arraystretch}{1.2} 
    \caption{Speedup Results of the partitioned CNN Models.}
    \begin{tabular}{p{1.4cm}p{0.8cm}p{0.8cm}p{0.8cm}p{0.8cm}p{0.8cm}p{0.8cm}}
        \hline
        Config &\multicolumn{3}{c}{\textbf{GPU Direct}} &\multicolumn{3}{c}{\textbf{GPU Indirect}}  \\
        \hline
         & Speedup Over DPU & Speedup over GPU & Layer Index & Speedup Over DPU & Speedup over GPU & Layer Index  \\
        \hline
        ResNet18 & 1.25x & 2.04x & 10 & 1.18x & 2.15x & 10 \\
        ResNet50 & 1.35x & 2.14x & 14 & 1.22x & 2.07x & 16  \\
        ResNet101 & 1.30x & 2.41x & 27 & 1.27x & 2.36x & 28 \\
        ResNet152 & 1.34x & 2.47x & 39 & 1.32x & 2.47x & 39 \\
        VGG16 & 2.48x & 1.4x & 12 & 2.33x & 1.34x & 17 \\
        MobileNetV2 & 1.1x & 3.15x & 16 & 1.1x & 3.3x & 16  \\
        \hline
        LeNet-5 & 1.01x & 1.15x & 4 & 1.01x & 1.28x & 5 \\
        \hline
    \end{tabular}
    \label{tab:wide}
\end{table}

\subsection{Split CNN Inference Results}
\label{sec:splitres}

\begin{figure*}[tbh!]
  \centering
  \includegraphics[width=0.8\linewidth]{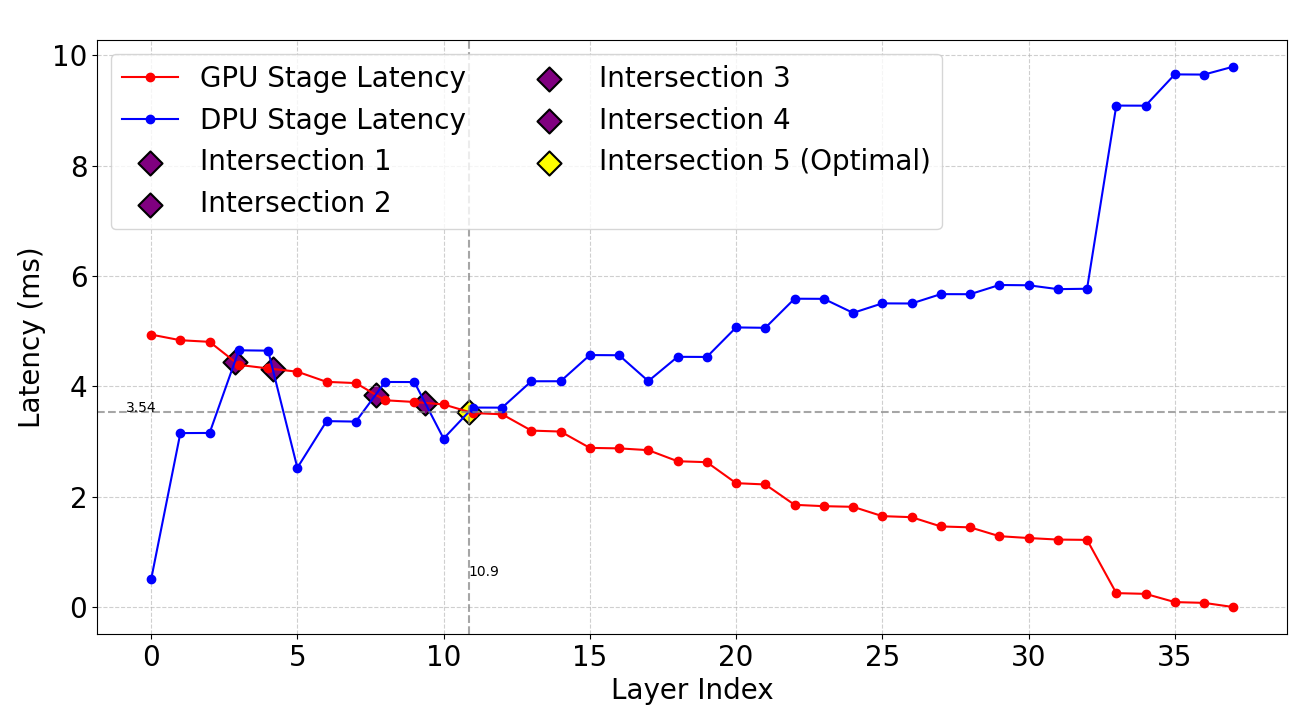}
  \caption{Latency of VGG16 inference across different partitions.}
  \label{vgg}
\end{figure*}

Table \ref{tab:wide} shows all speedup results. Remember that LeNet is only included to represent a pessimistic scenario due to the small number of layers in the CNN. Figure \ref{vgg} shows where the latency curves of each device intersect as the partition index increases. We can see that the partition index that gives the best runtime is when the first-stage latency is similar to the second-stage latency.  However, it is not a necessary condition. 

We obtain the biggest speedups over DPU with VGG16. That is because VGG16 execution is slower on the DPU than on the GPU, unlike the ResNet models. The VGG model includes a large number of dense 3x3 convolutions. These convolutions run in parallel on the GPU efficiently. ResNet models and MobileNetv2 include depthwise operations and residual connections, which are more efficient on the DPU. This yields path divergence and synchronization overheads on the GPU. On the DPU, data flows through AI Engine tiles efficiently without dynamic branching. Moreover, the storage of the residuals is also less efficient on the GPU. This causes cache evictions and forces the use of L2 cache. However, for the DPU, the output tensors that will be used in the following layers can be mapped to the on-chip SRAM of the DPU tiles, avoiding off-chip memory traffic. 

The experiments also show that increasing the layer count of the model leads to greater speedup. There are two reasons. First, deeper models have more split points, which makes them finer-grained; thus, we can balance the latency between the two stages of the pipeline better. The second reason is that when the model is short, the DPU-GPU data transfer time constitutes a big portion of the total runtime. This effect is prominent in the pessimistic scenario of the LeNet-5 model since its DPU-GPU transfer times make up nearly half of the DPU execution times. Finally, the results show that the GPU Indirect configuration decreases speedups due to the contribution of data transfer time to the second pipeline stage.

\begin{figure*}[tbh!]
  \centering
  \includegraphics[width=0.8\linewidth]{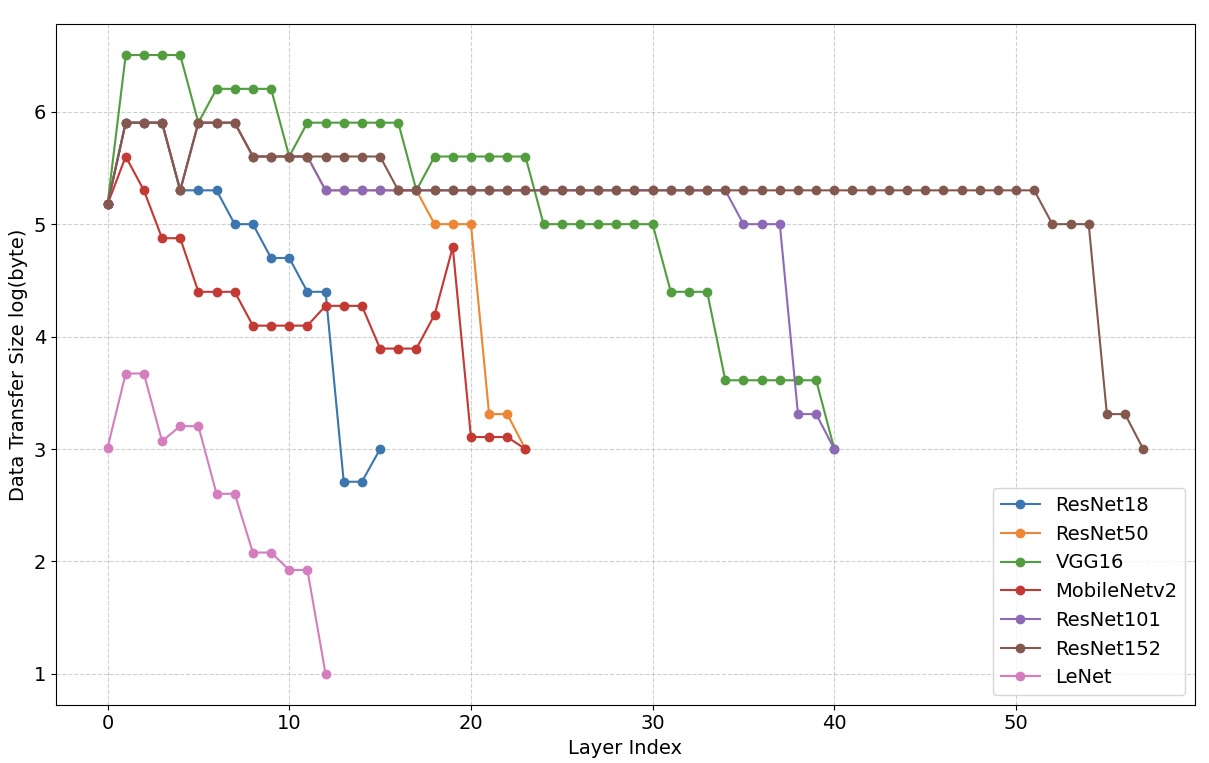}
  \caption{Exploration of data size transferred between DPU and host DRAM across partitions.}
  \label{data_size}
\end{figure*}

To sum up, the experiments show that splitting the model and pipelining it between DPU and GPU can reduce the latency. For each CNN model, the partition point that gives the maximum speedup changes according to device and model architectures.

\subsection{GNN-Based Partition Prediction}

\begin{table}[tb!]
\centering
    \renewcommand{\arraystretch}{1} 
    \caption{GNN Input Features and Random CNN Generation Space}
\begin{tabular}{lll}
\hline
\textbf{Feature Name}       & \textbf{Possible Values}                                                                                         & \begin{tabular}[c]{@{}l@{}}\textbf{Normalization} \\ \textbf{Constant}\end{tabular} \\ \hline
Layer Type          & \begin{tabular}[c]{@{}l@{}}[Conv, ReLU, Linear, \\ Average Pooling, \\ Max Pooling, Flatten]\end{tabular} & None                                                             \\ 
Kernel Size         & {[}1,3,7{]}                                                                                             & 7                                                                \\ 
Input Height        & {[}1,224{]}                                                                                             & 224                                                              \\ 
Input Width         & {[}1,224{]}                                                                                             & 224                                                              \\ 
Output Channels     & {[}32,64,128,256,512,1024,2048{]}                                                                       & 2048                                                             \\ 
Output Dimensions        & {[}128,256,512,1024,2048,4096{]}                                                                        & 4096                                                             \\ 
Input Dimensions         & \begin{tabular}[c]{@{}l@{}}{[}32,64,128,256,512,\\ 1024,2048,4096{]}\end{tabular}                       & 4096                                                             \\ 
Positional Encoding & {[}0,\#layers{]}                                                                                        & (\#layers-1)                                                     \\ \hline
\end{tabular}
\label{features}
\end{table}

We create a dataset consisting of 1700 random CNN models. These models are generated by sampling layer types and layer parameters from the ranges shown in Table~\ref{features}, while preserving tensor compatibility between consecutive layers. We then extract features for each CNN layer, such as layer type and kernel size, and convert each model into the graph representation described in Section \ref{sec:algorithm}. While doing that, the layer type is one-hot encoded, and the other features are normalized with their maximum values. The ground-truth latencies for each partition are obtained using the experimental methodology described in Section \ref{sec:meas}, and the ground-truth optimal partition index is selected as the partition that minimizes the resulting pipelined inference latency. Using these latency values and optimal split indices, we construct datasets for the latency-prediction and partition-index-prediction models, respectively. 

We trained two latency prediction models and an index prediction model for 200 epochs. Then, we evaluate using four metrics. \textit{Layer-wise accuracy} is the fraction of layers assigned to the correct partition across the dataset. \textit{Partition-index MAPE} is the absolute difference between predicted and ground-truth partition indices, normalised by model size and averaged over all models. \textit{Latency MAPE} is computed analogously on the latency of the resulting partition. Finally, \textit{Agreement Percentage} measures the percentage of models for which the predicted and ground-truth splits fall on the same side of the DPU-only/GPU-only baseline. Either both splits are better than running on DPU or GPU alone, or both are worse.

According to the results in Table~\ref{gnnres}, the end-to-end partitioning model presented more accurate results. On the other hand, it needs to be trained for each device pair, whereas the latency model needs to be trained for each device. Also, the model of \cite{chai} has a better MAPE. The reason is that their dataset is larger and the model has a larger capacity. Note that our model predicts a latency output for each node, which is more difficult to reason for deep learning models.

\begin{table}[tb!]

\centering
    \renewcommand{\arraystretch}{1} 
    \caption{GNN Evaluation Results}
\begin{tabular}{p{3.5cm}p{1.2cm}p{1.2cm}p{1.2cm}}
\hline
\textbf{\begin{tabular}[c]{@{}c@{}}Metric \\ Name\end{tabular}}         & \textbf{\begin{tabular}[c]{@{}c@{}}Latency \\ Model\end{tabular}} & \textbf{\begin{tabular}[c]{@{}c@{}}Partitioning \\ Model\end{tabular}} & \textbf{PerfSAGE \cite{chai}}                                            \\ \hline
Layer-wise Accuracy (\%)                                                 & 95.15                                                                       & 96.27                                                                            & N/A                                                          \\ 
\begin{tabular}[c]{@{}c@{}}Partition Index MAPE (\%)\end{tabular} & 4.73                                                                        & 3.93                                                                             & N/A                                                          \\ 
\begin{tabular}[c]{@{}c@{}}Latency MAPE (\%)\end{tabular}         & 8.63                                                                        & 7.52                                                                             & \textless{}5                                                 \\ 
\begin{tabular}[c]{@{}c@{}}Agreement Percentage (\%)\end{tabular}    & 86.73                                                                       & 95.92                                                                            & N/A                                                          \\ 
Dataset Size                                                            & 1700                                                                        & 1100                                                                             & 134912                                                       \\ \hline
Model Output                                                            & \begin{tabular}[c]{@{}c@{}}Layerwise \\ Latency\end{tabular}                & \begin{tabular}[c]{@{}c@{}}Partition \\ Indices\end{tabular}                     & \begin{tabular}[c]{@{}c@{}}Modelwise \\ Latency\end{tabular} \\ \hline
\end{tabular}
\label{gnnres}
\end{table}

\begin{figure*}[tbh!]
	\centering
	\includegraphics[width=0.8\textwidth]{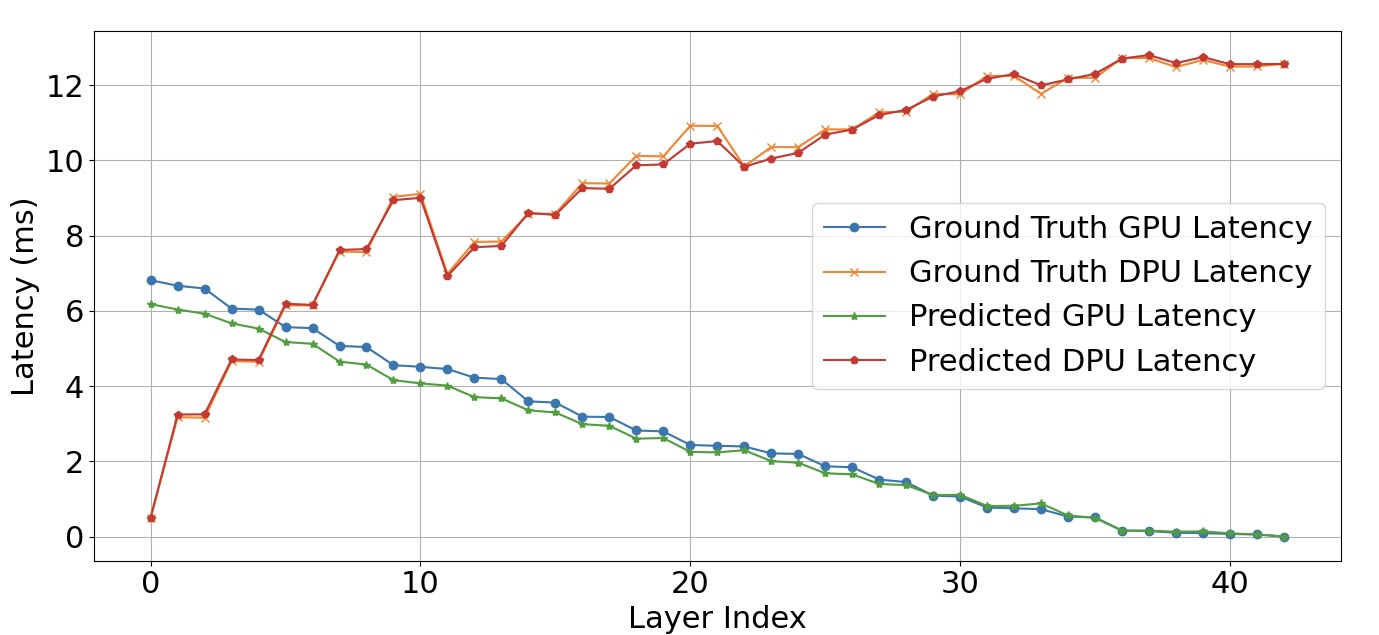}	
	\caption{Ground truth and predicted latency results for a 42-layer CNN model from the test set.}
	\label{custommodel}
\end{figure*}

\begin{figure*}[tbh!]
	\centering
	\includegraphics[width=0.8\textwidth]{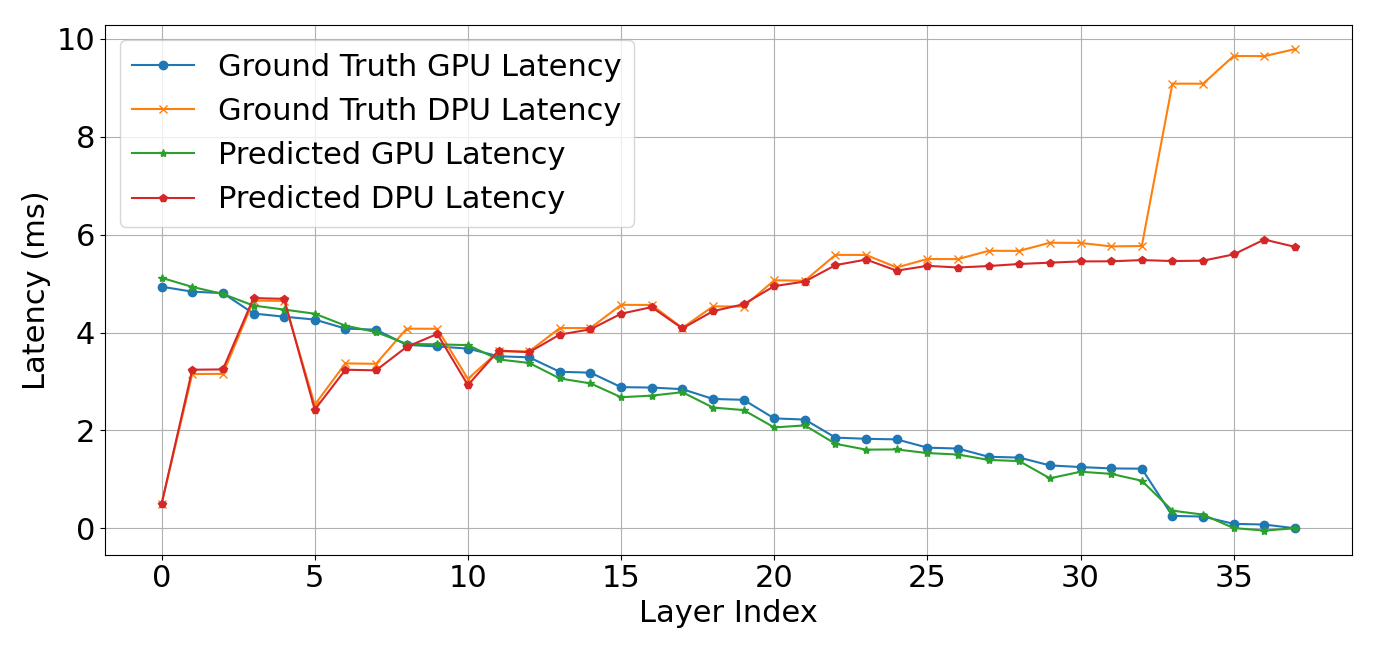}	
	\caption{Ground truth and predicted latency results for the VGG16 model.}
	\label{vgg16}
\end{figure*}

\section{Conclusions}
\label{sec:conclusions}

This paper has introduced Split CNN Inference, a novel approach that partitions CNN inference across DPU and GPU to reduce latency in image-streaming applications. The DPU processes the first partition near the source of data, and thus reduces the amount of data that needs to be sent to the GPU. The GPU processes the second partition. Split inference significantly improves efficiency, achieving up to 3.37× speedup over GPU-only and 2.48× over DPU-only execution across a myriad of well-known models, including LeNet-5, ResNet variants, VGG16, and MobileNetV2. 

The optimal partition points vary by model depth and hardware. Deeper networks generally offer greater gains when DPU and GPU performance are comparable. Thus, A GNN-based predictor has also been developed to identify suitable partition points, attaining near-optimal accuracy. This paper has also demonstrated different formulations of the Machine Learning problem and reported the results using non-NN approaches. 

Split CNN Inference has been tested on CNNs that have been optimised for execution on a single device. Split CNN Inference opens the window where Neural Architectural Search for CNNs considers from the start the two devices during training and tries to minimize the data transfer between DPU-GPU, which could produce even further reductions in latency.

\section*{Acknowledgments}
This work is partially funded by EPSRC EP/N035127/1 (LAMBDA project) and
EP/T026995/1 (EnnCore project). Mikel Luj\'an is supported by a Royal Society Wolfson Fellowship and an Arm/RAEng Research Chair Award.

\printbibliography

\end{document}